# Observation of spin Seebeck contribution to the transverse thermopower in Ni-Pt and MnBi-Au bulk nanocomposites

Stephen R. Boona[*,1], Koen Vandaele[2], Isabel N. Boona[3,4], David W. McComb[3,4], and Joseph P. Heremans[1,4,5]

[*]Corresponding author (boona.1@osu.edu)
[1]Department of Mechanical and Aerospace Engineering, The Ohio State University, USA
[2]Department of Inorganic and Physical Chemistry, Ghent University, Gent, Belgium
[3]Center for Electron Microscopy and Analysis, The Ohio State University, USA
[4]Department of Materials Science and Engineering, The Ohio State University, USA
[5]Department of Physics, The Ohio State University, USA

**Transverse thermoelectric devices produce electric fields perpendicular to an incident heat flux. Classically, this process is driven by the Nernst effect in bulk solids, wherein a magnetic field generates a Lorentz force on thermally excited electrons. The spin Seebeck effect (SSE) also produces magnetization-dependent transverse electric fields. SSE is traditionally observed in thin metallic films deposited on electrically insulating ferromagnets, but the films' high resistance limits thermoelectric conversion efficiency. Combining Nernst and SSE in bulk materials would enable devices with simultaneously large transverse thermopower and low electrical resistance. Here we demonstrate experimentally this is possible in composites of conducting ferromagnets (Ni or MnBi) containing metallic nanoparticles with strong spin-orbit interactions (Pt or Au). These materials display positive shifts in transverse thermopower attributable to inverse spin Hall electric fields in the nanoparticles. This more than doubles the power output of the Ni-Pt materials, establishing proof-of-principle that SSE persists in bulk nanocomposites.**

## Introduction

The spin Seebeck effect[1] (SSE) offers an alternate approach to conventional thermoelectric effects for solid-state heat-to-electricity energy conversion[2,3,4,5]. SSE involves the injection of heat into a polarized magnetic material, resulting in a thermal spin current. This spin current is typically detected by injecting it via spin pumping[6] into an adjacent thin film comprised of a non-magnetic metal with strong spin-orbit interactions, such as Pt. There, the inverse spin Hall effect (ISHE)[7] produces a transverse electric field proportional to the polarization and density of the incident spin current. These parameters are controlled experimentally by applying a magnetic field to alter the magnetic polarization, and by adjusting the temperature drop across the structure to alter the spin current density.

The geometry of SSE structures is identical to that in which the anomalous Nernst effect (ANE) is observed, except ANE occurs entirely within single-phase conducting ferromagnets. This is why metallic ferromagnets are avoided in experiments designed to isolate and understand the physics of SSE. However, combining SSE and ANE is a useful engineering strategy for enhancing the total transverse thermopower $S_{xyz}$ in ferromagnetic heterostructures without the need for large applied magnetic fields. Direct implementation of this concept for energy conversion means utilizing devices where the electrically active regions are thin films, which leads to a different set of problems: While thin films allow for the design of devices with mechanical flexibility[2], their small cross sectional area and inherently large electrical resistance limits the maximum possible power output of such devices to the mW level at most.

Here we demonstrate that SSE and ANE can be combined in an alternate bulk geometry better suited for energy conversion. Devices based on this concept hold potential for producing power at the W to kW level, since they are comprised of bulk materials that can be manufactured using large-scale production processes similar to those already in use for conventional thermoelectric devices. In this geometry, the entire device volume is both electrically and thermally active, enabling significantly lower source impedance and potentially higher voltage output compared to thin-film structures. Proof-of-concept is demonstrated here in a Ni-Pt composite, where the addition of Pt nanoparticles results in a significant enhancement of the transverse thermopower due to combined contributions from both the Nernst effect ($S_{\text{Nernst}}$) and SSE ($S_{\text{SSE}}$). Together with low electrical resistivity $\rho$, the transverse power factor $PF_{xyz} = S_{xyz}^2/\rho$ in the composites is increased 2-5 times compared to the reference sample, and the transverse thermoelectric figure of merit $zT_{xyz} = T \cdot PF_{xyz}/\kappa$ ($\kappa$ being the thermal conductivity and $T$ the absolute temperature) increases by an order of magnitude. This result establishes not only that SSE can be observed in bulk nanocomposites, but that it can also be utilized to produce significantly more electrical power from the same temperature gradient relative to single phase magnetic materials.

**Results**

**The spin Seebeck effect in thin-film structures**

A typical thin-film longitudinal SSE geometry is depicted schematically in Fig. 1(a). A temperature gradient $\nabla_x T$ is applied to a ferromagnetic insulator (FMI) possessing magnetization $M_z$ set by applied field $H_z$. $\nabla_x T$ generates a magnon flux, which carries heat[8] and spin along $x$. At the surface these magnons impinge on an adjacent normal metal (NM) film whose thickness is comparable to its spin diffusion length $l_s$, typically 1-10 nm. This spectrally dependent[9,10] spin pumping action polarizes free electrons in the NM, resulting in a perpendicular electric field $E_y$ via ISHE. The magnitude of $E_y$ depends on the polarization and magnitude of the injected spin current. This figure makes it apparent that from the perspective of energy conversion, electrical detection of thermally generated spin currents makes SSE functionally equivalent to a transverse thermopower.

From Fig. 1a, we define the SSE coefficient $S_{SSE} = |E_y/\nabla_x T|_{H_z}$. Typical values of $S_{SSE}$ in $Y_3Fe_5O_{12}$(YIG)/Pt heterostructures are approximately 0.1 µV K$^{-1}$ at 300K. Recent work on $Fe_3O_4$/Pt multilayer thin-films[11], YIG/NiO/Pt heterostructures[12], and spin flopped antiferromagnets[13] show coefficients near 6 µV K$^{-1}$ at various temperatures. These values are much lower than the longitudinal thermopower of classical thermoelectric devices (200 µV K$^{-1}$ in tetradymites[14]), so the intrinsic figure of merit in thin-film SSE heterostructures[4,15] is many orders of magnitude lower than in traditional semiconductor-based thermoelectrics.

Since spin is not conserved, the spin diffusion length of spin-polarized electrons in the NM ($l_s$) and magnons in the FMI ($L_s$) limits the geometry, size, and materials suitable for any SSE-based structures. For example, ISHE produces an electric field **E**$_{ISHE}$ in the NM only if the NM thickness $t$ is approximately equal to $l_s$ (*i.e.*, when only 0.001 vol% or 10 ppm of the total device volume is electrically active). If $t$ is much greater than $l_s$, then **E**$_{ISHE}$ is short-circuited by electrons in the region of NM where spin current has decayed. Second, $L_s$ limits the active thermal volume of FMI from which spin is injected into the NM. A conservative estimate for $L_s$ in YIG at 300K is 10 µm[16,17,18], though some experiments suggest a length scale of approximately 250 nm[19,20] is relevant to SSE. Regardless, this length is significantly shorter than the macroscopic sample and substrate thickness $L$, which is typically 0.5 mm. Since $\nabla_x T$ in $S_{SSE}$ is determined by $\Delta T_x/L$ and not $\Delta T_x/L_S$, only 1% of $\Delta T_x$ participates in generating $E_y$; although thermal excitation produces larger spin current densities than resonant or electrical methods[21], only a small fraction of the thermal energy is actually converted into electrical energy.

Pure SSE experiments require that the ferromagnetic material be insulating so that spin is transported only by magnons and not free electrons. Otherwise, **E**$_{ISHE}$ in the NM would mix with ordinary and anomalous Nernst thermopower from the ferromagnet, to give a total transverse thermopower $S_{xyz} = S_{Nernst} + S_{SSE}$ [22]. Careful separation of Nernst and SSE in bilayer structures is necessary when studying detailed origins of physical mechanisms[23]. For energy conversion, however, it is advantageous to combine Nernst and SSE into a single transverse thermopower, as long as both effects have the same polarity. This maximizes the voltage output per temperature gradient and enables lower electrical resistance, resulting in larger power output.

Multilayer structures are one way to accomplish this: Adding additional thermally and electrically active layers[11,24,25] can increase $S_{xyz}$ to 1.8 µV K$^{-1}$ and the extrinsic power factor $S_{xyz}^2/R$ ($R$ being the film sheet resistance) to 2 pW K$^{-2}$. Though these values can be reached at room temperature and with little to no applied magnetic field, the additional layers are still thin films. This means the electrically and thermally active regions of the device remain a Xsmall fraction of the total volume (much less than 1 vol%), so the devices have relatively high electrical impedance, and a significant portion of thermal energy goes unused. As an alternative approach, here we propose bulk composites as a scalable and robust way to make the entire device volume both electrically and thermally active, thus providing a viable pathway toward higher and more efficient power output.

**The spin Seebeck effect in bulk materials**

Fig. 1(b) illustrates the conceptual basis for SSE in a bulk composite geometry. In general, $\mathbf{E}_{ISHE} = \rho \tan \theta_{SH} (\mathbf{j}_S \times \boldsymbol{\sigma})$, where $\rho$ and $\theta_{SH}$ are the electrical resistivity and spin Hall angle of the NM, respectively, while $\mathbf{j}_s$ represents the spin current density in the NM, and $\boldsymbol{\sigma}$ indicates the spin current polarization. Experimentally, $\mathbf{j}_s$ is essentially proportional to $\nabla T$, and $\boldsymbol{\sigma}$ is determined by the ferromagnet magnetization tensor **M**, which is set by an external applied magnetic field **H**. As discussed above, $\mathbf{E}_{ISHE}$ is partially short-circuited by electrical conduction within the NM, resulting in attenuation that scales with $\exp[-t/l_s]$. Combining these and all other prefactors into a proportionality factor $A$, we have

$$\mathbf{E}_{ISHE} = A(\nabla T \times \mathbf{H}) \qquad (1).$$

To justify the composite geometry, we consider a spherical FMI particle coated with a NM film (Fig. 1(b)). When subjected to $\mathbf{H} = (0,0,H_z)$ and $\nabla T = (\nabla_x T,0,0)$, Eq. 1 indicates $\mathbf{E}_{ISHE}$ develops tangentially to the NM/FMI interface in the $(x,y)$ plane. Integrating $\mathbf{E}_{ISHE}$ over the entire sphere reveals $E_x = 0$ while $E_y(\varphi) = A|\nabla T||\mathbf{H}|\sin^2\varphi$, such that $E_y \geq 0$ for all $\varphi$. Assuming the sphere's radius $r \gg t$, then $S_{xyz}$ becomes:

$$E_y / \nabla_x T = A|\mathbf{H}| \int_0^{2\pi} \sin^2\varphi\, d\varphi = \pi A |\mathbf{H}| \qquad (2).$$

Equation (2) gives a value of $E_y$ that is ½ that of a slab with the same length as the periphery of the sphere. Admittedly, attenuation of $\mathbf{E}_{ISHE}$ via short-circuit through the NM is slightly enhanced in this spherical geometry over the traditional planar geometry, but we numerically estimate this factor to be approximately 15% for $t \approx l_s$.

By Eq. (2), we propose that any bulk composite material comprised of an FMI surrounded by a percolated network of sufficiently small NM grains should display a prominent $S_{xyz} = S_{SSE}$. Yet, this geometry is not the most favorable option for demonstrating such an effect, for two reasons. First, this approach requires compaction and sintering of FMI particles coated with thin shells of NM material. This presents processing difficulties, since a percolated nano-scale conducting pathway of NM grains must be maintained after sintering to observe SSE and extract electrical current. Second, spin transfer from FMIs into NMs depends on the FMI/NM interface quality[26], which is difficult to control in bulk materials.

We address both issues through the inverted geometry, wherein NM nanoparticles (NMNPs) are embedded within a ferromagnetic conductor (FMC) matrix. By the same logic described above, individual NMNPs in contact with FMC grains will display within them nonzero $\mathbf{E}_{ISHE}$, so long as the NMNPs' depth along $x$ is comparable to $l_s$. Current can then be extracted through the FMC matrix, and $\mathbf{E}_{ISHE}$ in the NMNP ($S_{SSE}$) will be superimposed on the background Nernst signal of the FMC ($S_{Nernst}$), as in Ref 22. This approach circumvents the need for a percolated conducting path of NM, enables the NMNP's to add their $\mathbf{E}_{ISHE}$ to the transverse voltage (if the polarities are chosen right), and simultaneously decreases the voltage source impedance. The latter effects both improve power efficiency.

This approach furthermore enables improved interfacial spin transfer efficiency, since spin currents entering the NMNPs arise from both magnons and spin-polarized electrons. However, the magnitude of $\mathbf{E}_{ISHE}$ is still limited by the degree of spin polarization in the FMC and the frequency of electronic spin-flip scattering at the FMC/NM boundaries. The sign of $\mathbf{E}_{ISHE}$ within the NM depends on the sign of $\theta_{SH}$ and the relative orientations of $\nabla T$ and $\mathbf{M}$, so we nominally expect the $S_{SSE}$ polarity to be independent of $S_{Nernst}$. Also, the electrical conductances of the FMC and the NM must be impedance matched to have the appropriate additive effect on the ISHE and Nernst voltages, which are electrically connected like voltage sources in parallel; *i.e.*, the FMC cannot have high electrical conductance.

**Results**

Details of the synthesis, measurement, characterization, and data analysis are provided in the Methods and Supplemental Information.

We synthesized five samples, summarized in Table 1. The first sample consisted of pure polycrystalline Ni nanoparticles (NiNP) produced via precipitation reaction. This sample was used to establish a baseline for $S_{Nernst}$ in the matrix material. Two other samples were NiNP-PtNP composites made via co-precipitation from solution of Ni and 2 wt% Pt. The fourth and fifth samples were made with MnBi as the host material and Au nanoparticles (AuNP) as the embedded NM phase.

In order to mitigate agglomeration of the PtNP and prevent possible alloying of Ni with Pt, the Ni-based composite samples were sintered at 250°C, well below the melting point of the constituent phases. This resulted in samples that were mechanically stable but only approximately 50% dense, with no lattice parameter shifts or other signs of alloying (see Supplementary Figure 1 and Supplementary Notes).

Fig. 2(a) shows a representative electron micrograph of a composite sample obtained after sintering. From this image, we see that the sample consists of Ni particles approximately 100 nm in diameter and partially coated in a thin layer of NiO approximately 5 nm thick. These Ni particles also appear to be decorated with small Pt nanoparticles approximately 5 nm in size (see Supplementary Figures 2 and 3 and Supplementary Notes for additional images). The PtNP sometimes form larger agglomerates nearly 20nm in size, as seen in parts of the image. These particles, including the agglomerates, are well within the size range where we expect an appreciable ISHE field to arise within the Pt due to thermal spin injection from the adjacent Ni.

The thin layer of NiO observed at the edges of the Ni grains implies that the transport properties of the samples we report here are likely much different from those of pure, dense, bulk Ni. This emphasizes the importance of utilizing a reference sample prepared in the same manner as the composites. Since it remains unclear how many interfaces in the sample are Ni/Pt and how many are Ni/NiO/Pt, we must consider what happens in the latter case. The presence of thin layers of electrically insulating NiO is certain to affect charge and spin transport between the Ni and Pt particles. However, since NiO is antiferromagnetic, spin currents can still propagate through thin layers of this material; this has been demonstrated in several recent studies, one of which observed spin currents transmitted through NiO layers up to 100 nm thick in Y$_3$Fe$_5$O$_{12}$/NiO/Pt heterostructures, as well as multiple studies that suggest the addition of NiO actually enhances the transmission of spin currents under certain conditions[12,27,28,29]. The presence of NiO undoubtedly has a detrimental effect on the electrical resistivity and power factor of both the reference and composite materials, though we anticipate these effects may be partially negated by a corresponding decrease in thermal conductivity.

With this in mind, we examine the transverse thermopower data collected at 125K and shown in Fig. 2(b) for the NiNP reference and NiNP-PtNP(1). The inset shows the difference in thermopower between the two samples overlaid with the NiNP magnetization at the same temperature. From the main figure, we see that the slopes of the transverse thermopowers vs field are approximately equal at fields larger than 0.5T, where the magnetization is saturated (the Ordinary Nernst Effect, ONE), while they are significantly different at fields smaller than 0.5T, where the magnetization is changing (the Anomalous Nernst Effect, ANE). The difference between the thermopowers of the two materials tracks the NiNP magnetization, consistent with the hypothesis that an ISHE field arises within the PtNP. We note that essentially no hysteresis was observed in the thermopower, and very minimal hysteresis was detected in the NiNP magnetization.

The ANE coefficients $[E_y/\nabla_x T]/H_z$ (*i.e.*, low-field thermopower slopes) for each of the NiNP-based samples are plotted vs temperature in Fig. 2(c). At every temperature, the samples containing PtNP have larger ANE coefficients than the sample without, indicating an additional magnetization-dependent contribution to the thermopower. A plot of the high field ONE coefficients vs temperature is included in Supplementary Figure 4. We do not observe any appreciable difference in ONE coefficient among the three samples, indicating that the addition of a few wt% PtNP has no significant or systematic effect on the transverse thermopower when the magnetization is saturated.

The electrical resistivity of all three samples is shown in Fig. 2(d). The samples containing PtNP are very similar to one another and slightly more resistive than the reference sample, which is approximately 10 times more resistive than dense Ni. The resistivity and ANE data are combined in Fig. 2(e), where we plot the intrinsic transverse power factor $PF_{xyz} = (S_{xyz})^2/\rho$ at various temperatures for all three samples. Since there is essentially no remnant magnetization in these materials, $PF_{xyz}$ was calculated using the transverse thermopower at $H_z = 0.1$T, an applied magnetic field easily reached with permanent magnets. These data show that the addition of PtNP increases $PF_{xyz}$ between 2-5 times compared to the matrix material alone. Thermal conductivity ($\kappa$) data are also presented in Fig. 2(f). Nanostructuring dominates $\kappa$ in these samples, as indicated by the relatively low magnitude and extremely flat temperature dependence. Like $\rho$, $\kappa$ is reduced by about 20% between the NiNP and the NiNP-PtNP sample 1; however, the variation in $\kappa$ between the two Pt-containing samples is larger than the corresponding change in $\rho$, presumably due to variations in microstructure resulting from slight differences in processing conditions.

These data can be combined to calculate a transverse thermoelectric figure of merit $zT_{xyz} = T \cdot PF_{xyz}/\kappa$, where $T$ is the absolute temperature. The transverse $zT_{xyz}$ values (not shown) follow the trend in $PF_{xyz}$ amplified by their variation in $\kappa$. The $zT_{xyz}$ values remain small, as they are in all conventional metals and SSE structures. Regardless, we observe an order of magnitude increase in $zT_{xyz}$ at 100K between the composite and reference samples, a dramatic improvement that can be attributed entirely to the spin Seebeck effect.

To provide a further check on the composite concept, we examined two additional samples where the host material (MnBi) has an ANE coefficient with polarity opposite to that of Ni, but the NMNPs (Au) still have positive spin Hall angles. Au was used instead of Pt because it has a much longer spin diffusion length, which allows for relatively large Au particles (above 50 nm) to be present without completely short circuiting the ISHE electric field. The wet-chemistry used to prepare Ni and Pt NP's cannot be applied to MnBi, which is moisture sensitive, so a different approach was utilized (see Methods).

The transverse thermopower of pure MnBi and a MnBi-AuNP composite at 300K are depicted in Fig. 3. To our knowledge, these are the first published data of the thermomagnetic properties of MnBi. The presence of AuNP in the composite shifts $E_y/\nabla_x T$ towards a more positive value, consistent with positive $\theta_{SH}$ in Au, and therefore with the existence of an ISHE contribution from the AuNP. The difference between the transverse thermopowers of the two materials is plotted in the lower left inset, along with the magnetization of MnBi powder.

Admittedly, any shift in transverse thermopower where the magnitude decreases is difficult to uniquely attribute to an ISHE field in the NMNP, since this may also result from simpler mechanisms like short circuits of the ANE electric field across larger NM particles. However, in this case, we observe the difference curve follows the same general field dependence of the magnetization, including a small hysteresis loop (upper right inset of Fig. 3), indicating that the shift in thermopower is not simply a constant offset arising from a short circuit of the ANE field. The size of this hysteresis loop is not particularly meaningful, since (unlike thin-film samples) we do not expect the ISHE field to follow exactly the hysteresis loop of the entire sample; instead, the ISHE hysteresis loop should reflect the magnetization of individual MnBi particles from which spin currents are thermally injected into neighboring AuNP. The absolute magnitude of the difference (the SSE voltage) appears to be significantly larger in this sample than it is in the Ni-Pt composites, though the relative change compared to the reference sample is similar. This is probably a reflection of differences in the ratio of electrical resistivity between the FMC and NM phases; Pt and Ni are roughly within a factor of two of each other, whereas our MnBi sample is at least an order of magnitude more resistive than Au (see Supplementary Figure 5). In this regard, the FMC matrix phase is less likely to short circuit the SSE voltage in MnBi-AuNP than in NiNP-PtNP, leading to larger $S_{SSE}$.

**Discussion**

Put together, these data demonstrate that the spin Seebeck effect can be observed and exploited not only in thin-film structures, but now also in bulk composite materials. Though we discuss our results above in terms of the conventional thermoelectric power factor concept, which we parse into intrinsic and extrinsic when comparing composites with thin films, it is perhaps more intuitive to simply consider the total power output of hypothetical devices based on each concept. Since bulk composites allow for electrical power to be extracted through their entire volume and thus provide far less internal resistance than thin films, such devices would be capable of producing significantly more total power from the same temperature gradient.

Broadly speaking, the composite concept applies to a wide variety of material combinations, and it invites further experimental and theoretical optimization of spin transport and spin pumping in irregular and/or random geometries. Further developments in materials selection and processing may lead to substantial increases in thermoelectric performance of magnetic composites beyond the order of magnitude improvement we report here, including large remnant transverse thermopowers that can produce voltage even in the absence of applied magnetic fields. If so, this approach may enable the development of low-cost SSE-based devices that require no vacuum processing and have dimensions like those of conventional thermoelectric materials, providing access to high power thermal energy conversion applications.

**Methods**

**Material synthesis**

All samples used for transport property measurements were polycrystals compacted from powders by spark plasma sintering (SPS) into 10 mm diameter cylindrical pellets approximately 1-2 mm thick. The nanoparticles used in this study were either Au (AuNP), which had diameters before sintering of approximately 50 nm and were purchased commercially (Sigma Aldrich), or Pt (PtNP), which had diameters before sintering of 2-5 nm and were chemically and mechanically dispersed among Ni nanoparticles during synthesis at OSU.

PtNP and NiNP were synthesized by reducing $Ni(NO_3)_2 \cdot 6H_2O$ with $NaBH_4$ in the presence of $H_2PtCl_6 \cdot 6H_2O$ and sodium citrate as complexing agent. The resulting NiNP were amorphous (a-NiNP) and intimately mixed with PtNP, as observed by imaging in the scanning transmission electron microscope (STEM). The a-NiNP were then converted to crystalline NiNP by post reduction treatment with $N_2H_4 \cdot H_2O$. Hydrogen hexachloroplatinate(IV) hydrate, ca. 40% Pt, and sodium borohydride powder 99% were purchased from Acros Organics. Sodium citrate dihydrate >99%, nickel nitrate hexahydrate 99.99%, 64-65% hydrazine monohydrate and 99.8% ethylene glycol were supplied by Sigma Aldrich. All chemicals were used as received.

In a typical synthesis, 4 g $Ni(NO_3)_2 \cdot 6H_2O$ salt and 0.0428 g $H_2PtCl_6 \cdot 6H_2O$ were dissolved in 250 mL water, stabilized by 8.77 g sodium citrate. Subsequently, the solution was heated to 80°C and 1.56 g $NaBH_4$ was dissolved in 20 mL water. Then 7.15 g NaOH was added and the mixture was stirred vigorously for 15 min. Next, 10.8 mL $N_2H_4 \cdot H_2O$ in 20 mL ethyleneglycol was added to the solution and stirred for one hour at 80 ˚C. The nanocomposite powder was collected magnetically and washed thoroughly with water and ethanol. The dried powder was then compacted by spark plasma sintering under 50 MPa of uniaxial pressure at 250°C held for 30 minutes in a graphite die under continuous vacuum.

Polycrystalline powders of MnBi were synthesized through a combination of arc melting, grinding, sieving, and annealing, using the methods similar to those described in Ref 30. The MnBi-AuNP composite sample was created by mixing 5 vol% AuNP with 95 vol% MnBi powder (-125 mesh) and sonicating in a hexane bath at 35 kHz for 20 minutes. Hexane was used to prevent decomposition of MnBi, which is moderately air and water sensitive. The suspension was dried then sintered under 50 MPa of uniaxial pressure at a temperature of 200°C held for 5 minutes in a graphite die under continuous vacuum. The control sample of MnBi was synthesized from the same batch of powder under the same conditions but without AuNP.

**Structural and compositional characterization**

X-ray diffraction (XRD) analysis was performed using a Rigaku MiniFlex Diffractometer with Cu-K$_\alpha$ radiation. Based on the XRD results, we successfully synthesized high purity polycrystalline MnBi and MnBi-AuNP materials containing no discernible impurity phases (see Supplementary Figure 6 and Supplementary Notes). We also observe no discernible peak shift in the Ni-Pt composites, indicating no alloying occurred between Ni and Pt during sintering at 250°C.

Microstructural analysis was performed on several samples of control and composite materials by scanning electron microscopy (SEM) and scanning transmission electron microscopy (STEM) with high-angle annular dark field (HAADF) imaging and x-ray energy dispersive spectroscopy (XEDS). The HAADF-STEM imaging and XEDS spectrum collection was performed using a monochromated FEI Titan 60-300 STEM equipped with a Super-X XEDS collection system. The SEM work and XEDS spectra were collected using the FEI Sirion field emission SEM equipped with the EDAX Octane Super SDD system.

SEM analysis of the AuNP shows Au inclusions ranging in size from approximately 50 nm particles up to 20 micron agglomerates. STEM and XEDS analysis of the NiNP-PtNP pre-sintering indicated 2-5 nm PtNP. After sintering, we observed various nanostructured features that included 20 nm Pt agglomerates and what appear to be PtNP under 5nm in diameter widely dispersed and decorating the Ni grains. Additional representative images are provided in the supplemental information.

**Measurements**

Parallelepiped pieces of each sample were cut using a rotary diamond saw. Silver epoxy (EpoTech) was used as a conducting adhesive to attach gold coated copper leads, heat spreaders, and heat sinks to the samples in a five probe configuration. All transport measurements were performed in a Quantum Design Physical Property Measurement System (PPMS) via the Thermal Transport Option (TTO). Transverse thermopower measurements were performed using the five probe technique with a home-built breakout box, while electrical resistivity was measured using a conventional AC four probe method. Seebeck coefficient and thermal conductivity were determined using a continuous measurement protocol.

For transverse thermopower measurements, a steady state temperature difference $\Delta T_x$ was established and then the corresponding transverse voltage $\Delta V_y$ was recorded while continuously sweeping the applied magnetic field $H_z$ back and forth between maximum and minimum values of $H_z = 30$ kOe at a constant rate of 50 Oe/s. The data were analyzed by averaging out the odd components to isolate the hysteresis curve for each temperature point. These signals were then normalized by the corresponding field-dependent temperature difference as well as by the associated lengths in order to arrive at the intrinsic $S_{xyz}$ coefficient.

**Data Availability**

The data that support the findings of this study are available from the corresponding author upon request.

## Acknowledgements

This work is supported by the U. S. National Science Foundation MRSEC program under grant No.DMR1420451 and by the U. S. Army Research Office MURI program under grant No. W911NF-14-1-0016. Microscopy was performed at The Ohio State University (OSU) Center for Electron Microscopy and Analysis (CEMAS).

## Author contributions

SRB devised the study with feedback and supervision from JPH, who also derived the model equations. SRB also synthesized the MnBi powder, sintered the pellets, performed the transport and XRD measurements, and conducted the data analysis. KV performed the wet chemistry, including synthesis of the Ni-PtNP powders. INB performed all microscopy analysis with feedback and supervision from DWM. All authors contributed to writing and editing the manuscript and supplement.

## Additional information

Supplementary information is available online.

**Competing interests**

The authors declare no competing financial interests.

**Tables**

**Table 1. Summary of the samples studied here.**

[†]*Based on the spin Seebeck sign convention, which is opposite that of Gerlach.*

**Figures**

**Figure 1. Schematic representation of the spin Seebeck effect in different geometries.** Spatial directions ($x,y,z$) are specified by the legend at the top. The concentric circles indicate the applied magnetic field $H_z$ and magnetization $M_z$ for both panels are pointing out of plane. (a) Geometrical layout of a conventional spin-Seebeck effect (SSE) geometry showing a normal metal (NM) film (red) on top of a ferromagnetic insulator (FMI, green). A spin current $j_{S,x}$ is driven by a temperature gradient $\nabla_x T$ within the FM. The polarization direction of $j_{S,x}$ is determined by $H_z$ and $M_z$. This spin current is injected into the NM, where the inverse spin Hall effect (ISHE) produces an electric field $E_y$. (b) One quadrant of the same structure folded upon itself in an FMI (green) / NM (red) core-shell configuration. Under the same conditions as (a) ($H_z$, $M_z$, $\nabla_x T$), an ISHE field **E** again arises in the NM, but now its directional components must be taken into account. Integrating **E** over the full circle shows that $E_x$ cancels out, but non-zero $E_y$ remains (see Eq. 2). This very general idea can be applied to a variety of arbitrary interface geometries, thereby providing the conceptual basis for SSE in random bulk composites.

**Figure 2. Microstructure and transport properties of Ni-Pt composites.** (a) High-angle annular dark field (HAADF) image of NiNP-PtNP(2) overlaid with false-color elemental mapping of Pt (red), Ni (green), and O (blue) determined by x-ray energy dispersive spectroscopy (XEDS) in a scanning transmission electron microscope (STEM). (b) Transverse thermopower vs applied field at 125K for porous NiNP (red) and porous NiNP-PtNP(1) (blue). The slopes at low and high fields give the ANE and ONE coefficients, respectively. The inset shows the difference between the red and blue curves (purple), along with the sample magnetization (orange). (c) ANE coefficients of the three samples vs temperature. Solid lines are a guide for the eye. (d) Resistivity of all three samples vs temperature. (e) Temperature dependence of the intrinsic power factor of NiNP (red circles), NiNP-PtNP(1) (solid blue squares), and NiNP-PtNP(2) (empty blue squares) at $H_z = 0.1$T. Solid lines are a guide for the eye. (f) Thermal conductivity of all three samples vs temperature.

**Figure 3. Transverse thermopower of MnBi-Au composites at 307K.** The red trace indicates the thermopower of MnBi, while that of MnBi-AuNP is shown in blue. The lower left inset shows the difference between these curves (purple) as well as the magnetization of the MnBi at 307K (orange), while the upper right inset emphasizes the hysteresis loop in the difference plot.

**Tables**

| Sample name | Host compound | Nanoparticles | NP size (nm) | ANE of host[†] | $\theta_{SH}$ of NP |
|---|---|---|---|---|---|
| NiNP | Ni | Ni | 100 | + | N/A |
| NiNP-PtNP(1) | Ni | Ni + Pt | 100 (Ni), 5 (Pt) | + | + |
| NiNP-PtNP(2) | Ni | Ni + Pt | 100 (Ni), 5 (Pt) | + | + |
| MnBi | MnBi | N/A | N/A | - | N/A |
| MnBi-AuNP | MnBi | Au | 50 nm | - | + |

**Table 1. Summary of the samples studied here.**
[†]Based on the spin Seebeck sign convention, which is opposite that of Gerlach.

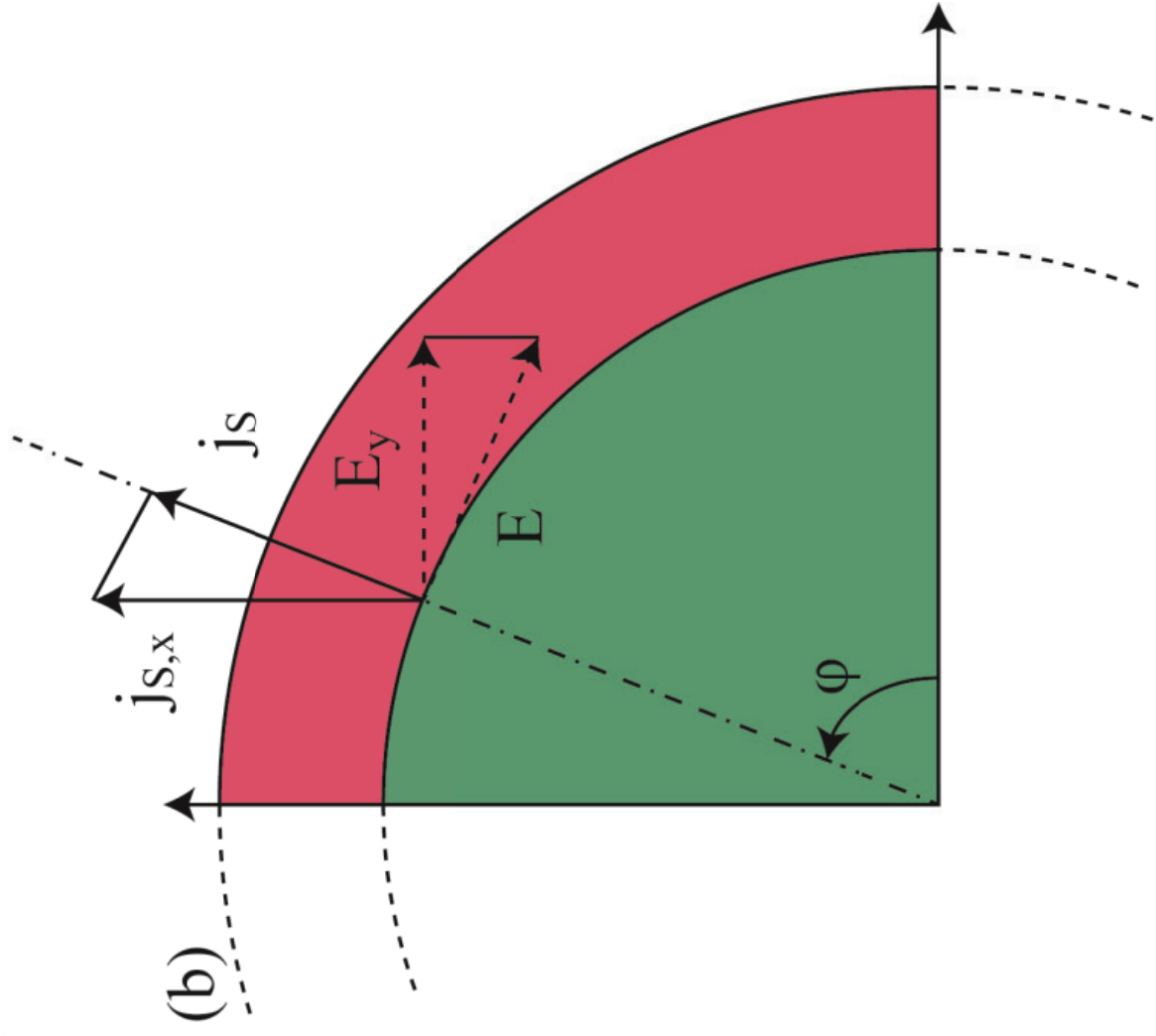

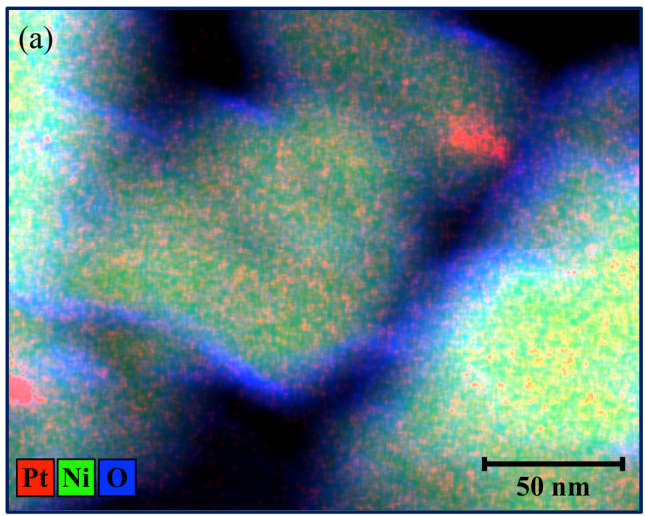
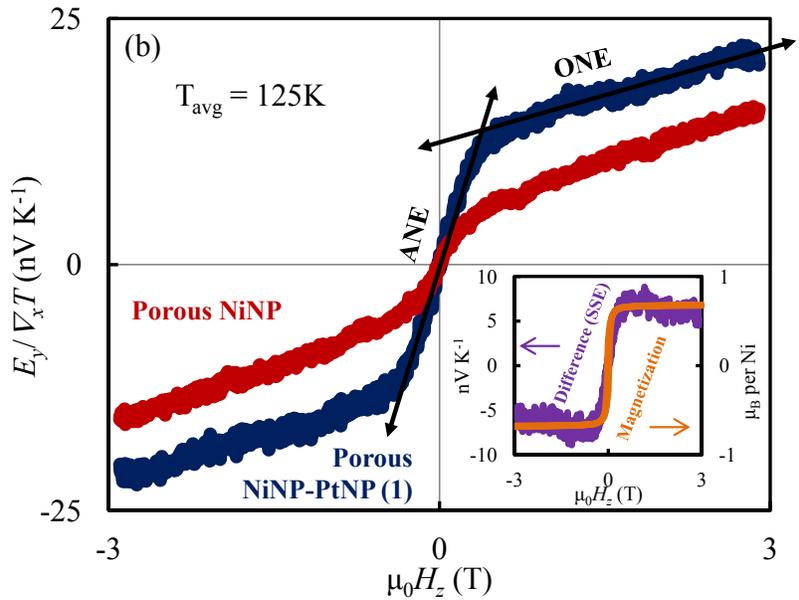
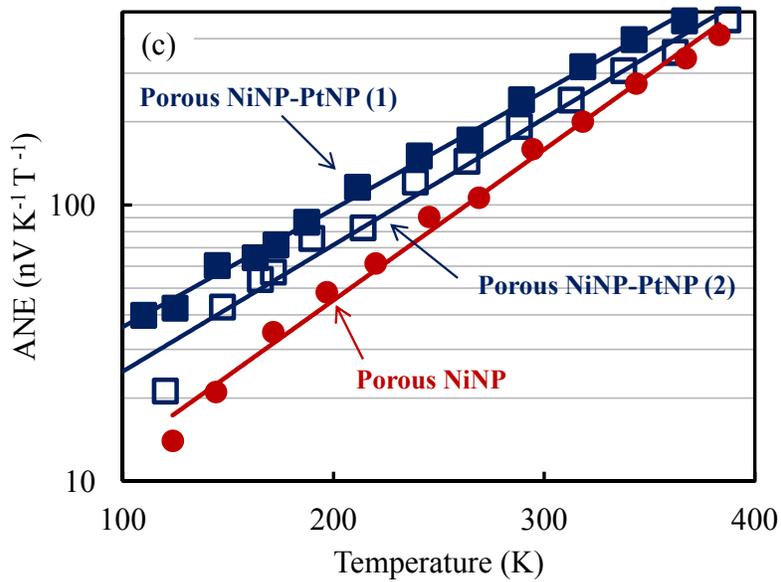
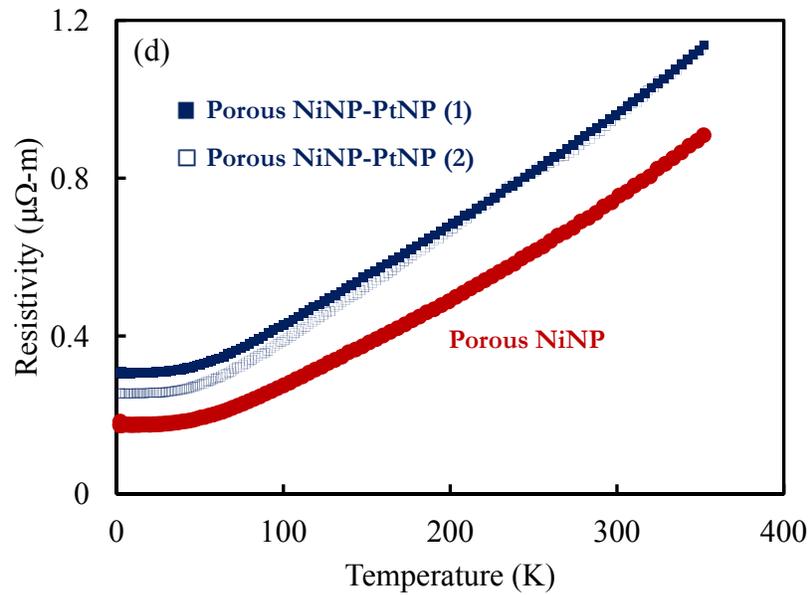
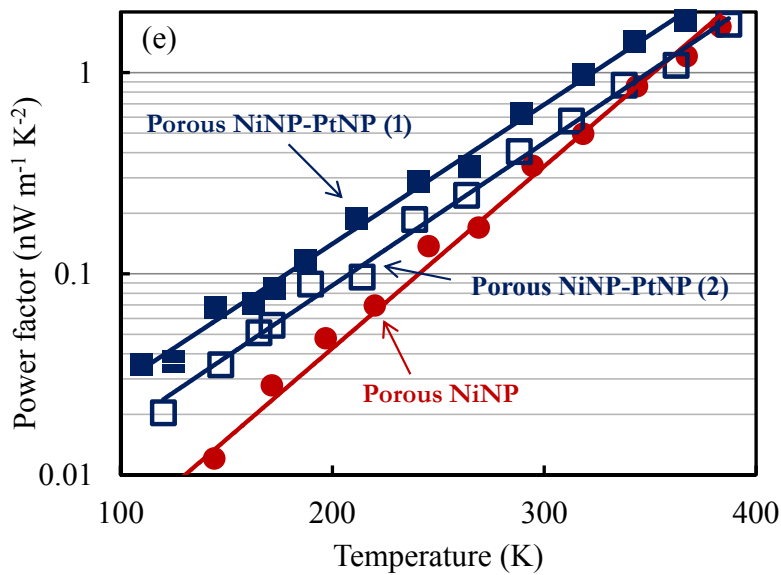
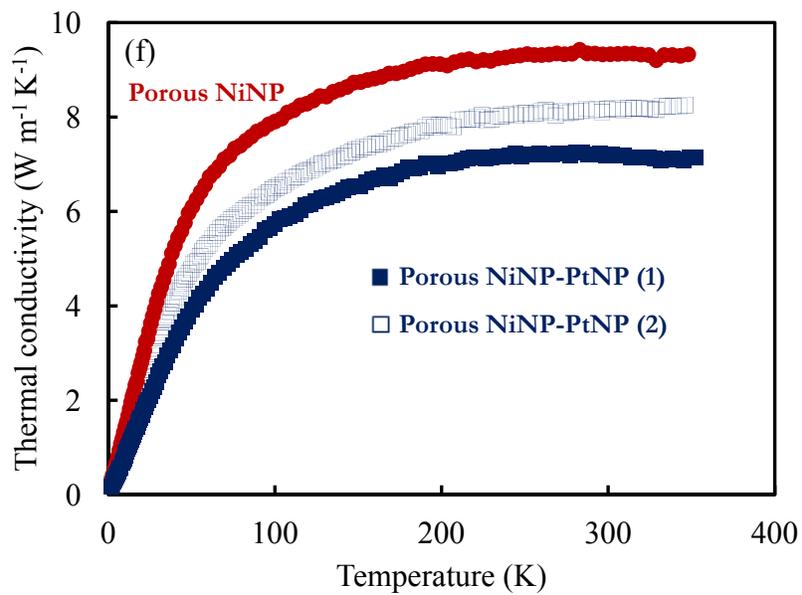

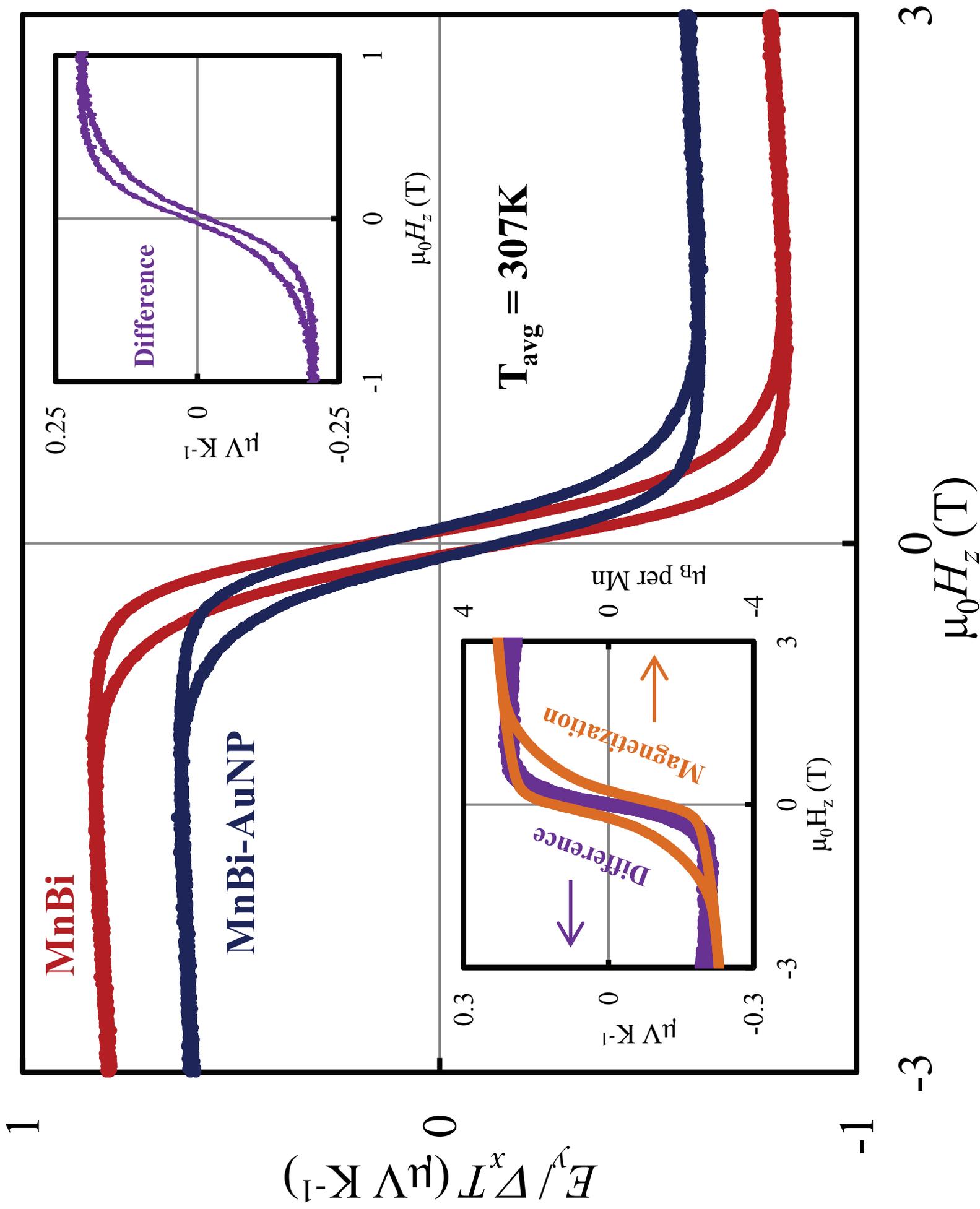

**Supplementary Figures**

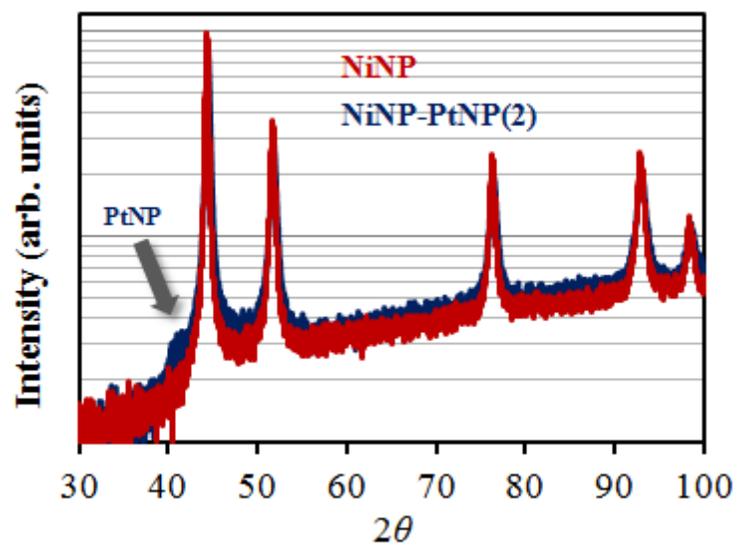

*Supplementary Figure 1. Powder x-ray diffractometry of sintered NiNP and NiNP-PtNP(2).*

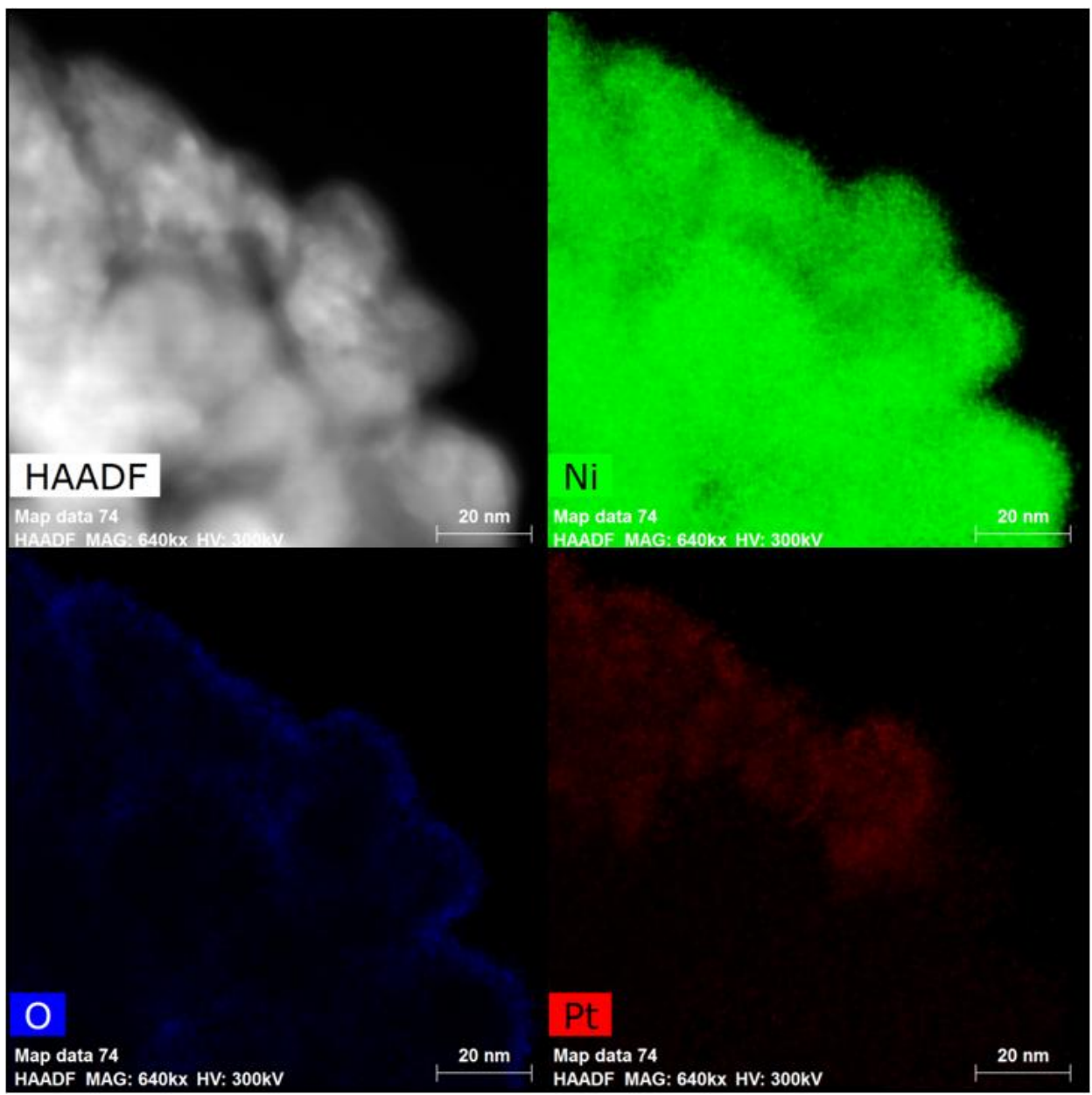

*Supplementary Figure 2. HAADF-STEM image and XEDS elemental maps of NiNP-PtNP(2). HAADF-STEM (top left) image is displayed, along with elemental maps of Ni (green, top right), O (blue, bottom left), and Pt (red, bottom right).*

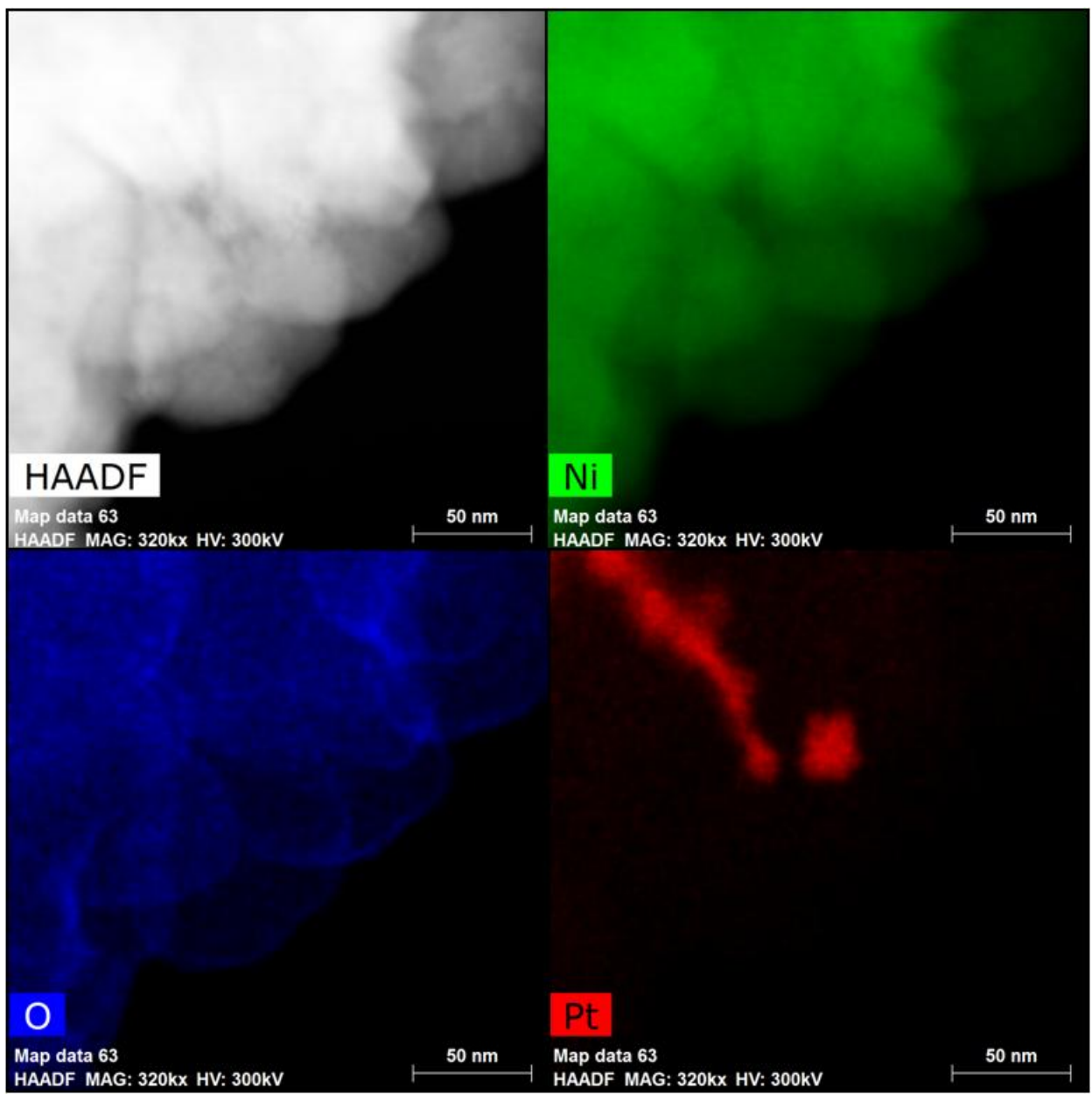

*Supplementary Figure 3. Additional HAADF-STEM XEDS elemental maps of NiNP-PtNP(2). HAADF-STEM (top left) image is displayed, along with elemental maps of Ni (green, top right), O (blue, bottom left), and Pt (red, bottom right).*

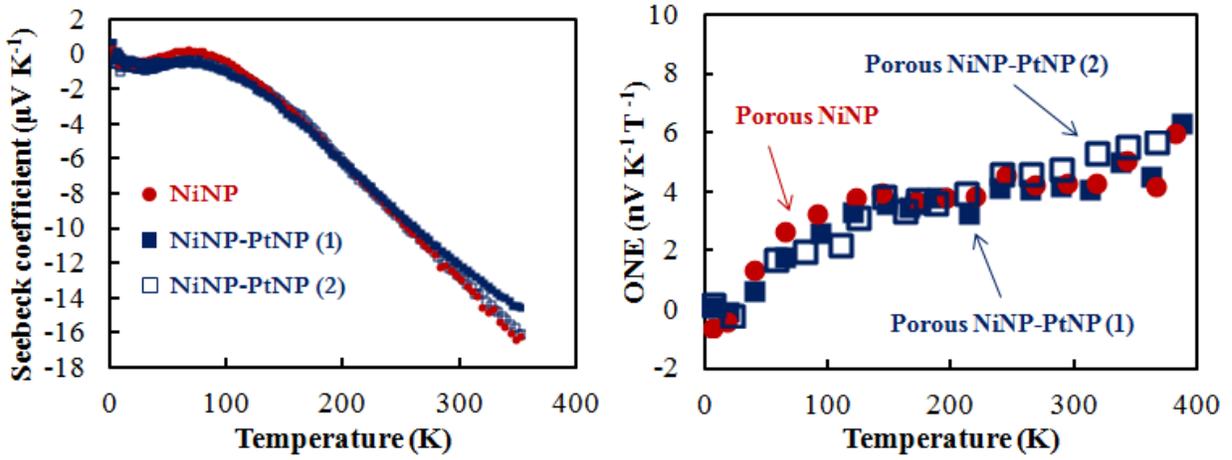

*Supplementary Figure 4. Seebeck coefficient (left panel) and ordinary Nernst coefficient (right panel) of the three Ni-based samples.*

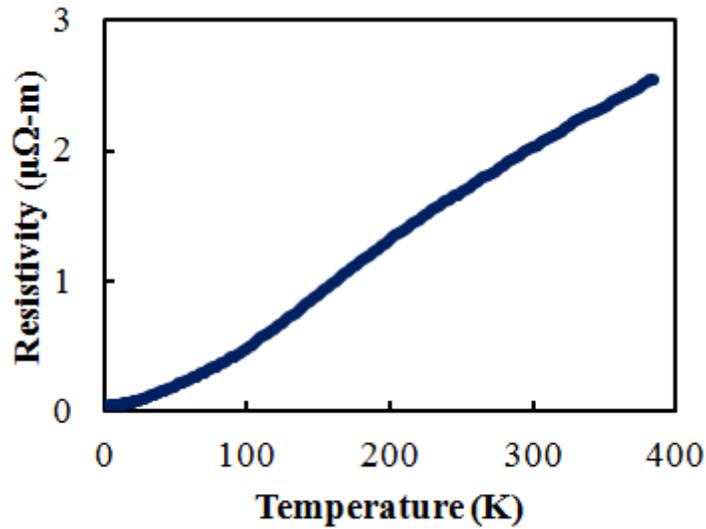

*Supplementary Figure 5. Electrical resistivity vs temperature for polycrystalline MnBi.*

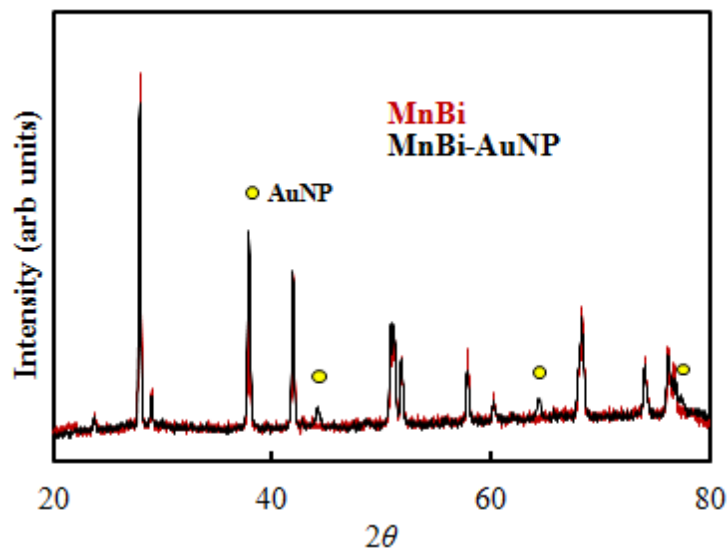

*Supplementary Figure 6. Powder XRD of sintered MnBi and MnBi-AuNP. No trace of a phase other than MnBi is seen, but the detection limit of secondary phases in XRD spectra is limited to a few %.*

**Supplementary Notes**

**Electron microscopy**

Supplementary Figure 2 shows a second representative HAADF-STEM image (first image is in the main text) of the NiNP-PtNP(2) sample with associated false-colored elemental maps of Ni, Pt, and O. This image shows what appear to be Ni particles less than 50nm in size coated in NiO, with a portion of the image coated in a thick layer of Pt particles and/or agglomerates ~20 nm in size.

A third representative HAADF-STEM image of NiNP-PtNP(2) is shown in Supplementary Figure 3, along with the associated false-colored elemental maps of Ni, Pt, and O. This image shows what appear to Ni particles less than 100nm in size coated in NiO and decorated with Pt particles. A concentrated vein of Pt particles and/or agglomerates ~10-20 nm thick is visible in the upper center/left of the image.

**X-ray diffractometry**

Supplementary Figure 1 shows the powder XRD results obtained for the sintered NiNP and NiNP-PtNP(2) samples. We observe no noticeable shift in lattice parameter after sintering, indicating no detectable alloying of Pt and Ni. Instead, near $2\theta = 40°$ we see the primary diffraction peak of pure Pt metal. All peaks are clearly broadened relative to their intensity, indicating sub-100 nm coherent scattering volumes.

Supplementary Figure 6 shows the powder x-ray diffraction (XRD) results obtained for the sintered MnBi and MnBi-AuNP samples made from the same batch of MnBi powder. As seen in the figure, we obtain high purity MnBi with no discernible impurity peaks visible in either sample. The only difference between the control and composite samples are the additional peaks of Au present in the latter. Similar to what is seen in Supplementary Figure 1, the Au peaks are clearly broadened relative to their intensity, indicative of sub-100nm coherent scattering volumes.